# Superconductivity and Equation of State of Lanthanum at Megabar Pressures

Wuhao Chen[1], Dmitrii V. Semenok[2], Ivan A. Troyan[3], Anna G. Ivanova[3], Xiaoli Huang[1,*], Artem R. Oganov[2,4], and Tian Cui[5, 1,*]

[1]State Key Laboratory of Superhard Materials, College of Physics, Jilin University, Changchun 130012, China

[2]Skolkovo Institute of Science and Technology, Skolkovo Innovation Center, 3 Nobel Street, Moscow 143026, Russia

[3]Shubnikov Institute of Crystallography, Federal Research Center Crystallography and Photonics, Russian Academy of Sciences, Moscow 119333, Russia

[4]International Center for Materials Discovery, Northwestern Polytechnical University, Xian 710072, China

[5]School of Physical Science and Technology, Ningbo University, Ningbo 315211, China

*Corresponding authors: huangxiaoli@jlu.edu.cn and cuitian@jlu.edu.cn

**Abstract**

Lanthanum (La) is the first member of the rare-earth series of elements that has recently raised considerable interest because of its unique La superhydride $LaH_{10}$ and its own superconducting properties. There has been a lack of experimental evidence for the equation of state (EoS) and superconductivity of La at pressures exceeding one megabar. Here, we extend the pressure range up to 140 GPa to explore the EoS and superconductivity of La via the electrical resistance and X-ray diffraction measurements. The Le Bail refinement of the experimental XRD patterns determined the phase transition sequences and discovered a new distorted *fcc*-La phase (space group *Fmmm*) above 78 GPa. All the experimental pressure-volume data were fitted by the third-order Birch-Murnaghan equation: $V_0$ = 35.2 (4) Å$^3$, $B_0$ = 27 (1) GPa and $B_0$' = 4. Superconductivity still existed in the newly discovered distorted *fcc*-La with an onset critical temperature $T_c$ of 9.6 K at 78 GPa, which decreases to 2.2 K at 140 GPa. We have extrapolated the upper critical magnetic field $\mu_0H_{c2}$ at 140 GPa, and the theoretical calculations predict the electron-phonon coupling mechanism with the superconductivity parameters consistent with the experimental results.

## Introduction

Since the discovery of superconductivity (SC) in 1911 [1], scientists have searched for materials that can conduct electricity without resistance below the superconducting transition temperature ($T_c$). So far, tremendous efforts have been devoted to exploring the high-$T_c$ SC in a variety of materials such as cuprates [2], iron-based superconductors [3], and hydrogen-rich compounds [4]. Among them, the SC in hydrides has been successfully realized with the discovery of novel compounds in the H-S system at high pressures with the $T_c$ of up to 203 K. The discovery of high-temperature superconductivity in $Im\bar{3}m$-H$_3$S by theoretical [5-6] and experimental [7-9] methods hints that even a room-temperature SC can be achieved in hydrogen-rich materials. Hence, the search for superconducting polyhydrides at very high pressures has brought about a new round of research upsurge in physics. Metal hydrides are interesting materials for realizing high-$T_c$ superconductivity when they form unusually high stoichiometric ratios. In particular, metal lanthanum (La) can react with hydrogen yielding $Fm\bar{3}m$-LaH$_{10}$, which has recently been reported as a high-$T_c$ superconductor with the critical temperature above 250 K at 170 GPa [10-11]. In the framework of the total experimental and theoretical investigation of La-H system, a study of structural and electronic properties of metal La at ultrahigh pressures is necessary for better understanding the roles of La and H atoms in the observed SC.

The unique physical and chemical properties of rare earth metals have attracted interest for decades. The mostly trivalent rare earth metals from La through Lu possess a similar *d*-electron character near Fermi energy, none of them superconducting at ambient pressure except metal La [12]. As the first member of the rare-earth series of elements, La can exist in both double hexagonal-close-packed (*dhcp*) phase and face-centered cubic (*fcc*) phase, while a *bcc* structure is favorable at high temperatures near the melting point [13-15]. Due to a high electronic density of states at Fermi surface and a specific phonon spectrum, one would expect a strong electron-phonon coupling, and, therefore, a reasonably high superconducting transition temperature for La. The *dhcp* and *fcc* phases yield the superconducting transition temperatures $T_c$ near 5 K and 6 K at ambient pressure, respectively [16]. Both structures show significant increase in $T_c$ with pressure ($dT_c/dP$ ~ 0.87-1 K/GPa): at 4 GPa it reaches ≈ 9.3 K, around 17 GPa — 13 K [17-18]. The *dhcp* phase undergoes several structural transformations, first transforming to the *fcc* structure near 2.2 GPa, then to a distorted *fcc* ($R\bar{3}m$) structure at about 5.4 GPa, then returning to the original *fcc* phase at 60 GPa [16, 17]. At the same time, the critical temperature demonstrates a complex behavior with several "waves" (broad maxima) and anomalies up to 50 GPa [17]. Until now, the equation of state (EoS) and superconductivity of metal La were not studied under pressures over 100 GPa.

In this work, we present the new data obtained by comprehensive studies of the structural and superconducting properties of La at pressures up to 140 GPa. A new distorted *fcc*-La phase was discovered with *Fmmm* symmetry above 78 GPa. Superconductivity survives in La with critical temperature $T_c$ of 2.2 K at 140 GPa. We have also extrapolated the upper critical magnetic field ~ 0.32-0.43 T upon the applied magnetic field at 140 GPa.

**Experimental and Theoretical Methods**

We used the target La sample purchased from Alfa Aesar, with the purity of 99.9%. The assembly used for the electrical resistance measurements is shown in Fig. 1. We used a piston-cylinder diamond anvil cell (DAC) made of a Be-Cu material, the diamonds had a culet of 100 μm in diameter beveled at 8° to a diameter of about 300 μm. The sample chamber consisted of a tungsten gasket with an $Al_2O_3$ insulating layer. The excess $Al_2O_3$ was also used as a pressure-transmitting media (PTM). The piston diamond was coated with four 1-μm-thick Mo electrodes connected to the external wires using a combination of 5-μm-thick Pt shoes soldered onto brass holders. The 5-μm-thick La sample was placed on the Mo electrodes and packed in with the aluminum oxide in an argon glove box. The pressure was determined from the Raman shift of the diamond anvil edge excited with a 532 nm laser [19].

We have performed in situ high-pressure XRD patterns of La sample in the pressure range of 7-140 GPa on synchrotron beamline 16-BMD at Advanced Photon Source using a focused monochromatic X-ray beam (λ = 0.434 Å). In experimental run 2, the synchrotron XRD patterns of the La sample in a pneumatic DAC with a 50-μm culet were recorded on the ID27 synchrotron beamline at the European Synchrotron Radiation Facility (Grenoble, France) with a focused monochromatic X-ray beam of 33 keV (λ = 0.3738 Å), in the pressure range of 107-148 GPa. In both experimental runs, MgO was used as PTM and pressure gauge [20]. $CeO_2$ standard was used for the calibration of the experimental parameters (sample-to-detector distance, detector's tilt angle, and the beam center). The experimental XRD images were analyzed and integrated using the Dioptas software package (version 0.4) [21].

The calculations of the superconducting transition temperature $T_c$ were carried out using the QUANTUM ESPRESSO (QE) package [22]. The phonon frequencies and electron-phonon coupling (EPC) coefficients were computed using density-functional perturbation theory [23], employing the plane-wave pseudopotential (PP) method, Perdew-Burke-Ernzerhof and Perdew-Zunger exchange-correlation functionals [24]. In our *ab initio* calculations of the electron-phonon coupling (EPC) parameter $\lambda$, the first Brillouin zone was sampled using 2×2×2 and 4×4×4 q-points meshes, and a denser 24×24×24 k-points mesh (with Gaussian smearing and $\sigma = 0.025$ Ry, which approximates the zero-width limit in the calculation of $\lambda$). The critical temperature $T_c$ was calculated using the Allen-Dynes formula [25]. The EoS for a distorted *fcc*-La was calculated by performing structure relaxations of cell parameters at various pressures using density-functional theory (DFT) within the generalized gradient approximation and using the projector-augmented wave (PAW) method as implemented in the VASP code [26]. The plane wave kinetic energy cutoff was set to 600 eV, while the Brillouin zone was sampled using Γ-centered k-points meshes with the resolution of $2\pi \times 0.05$ Å$^{-1}$.

## Results and Discussion

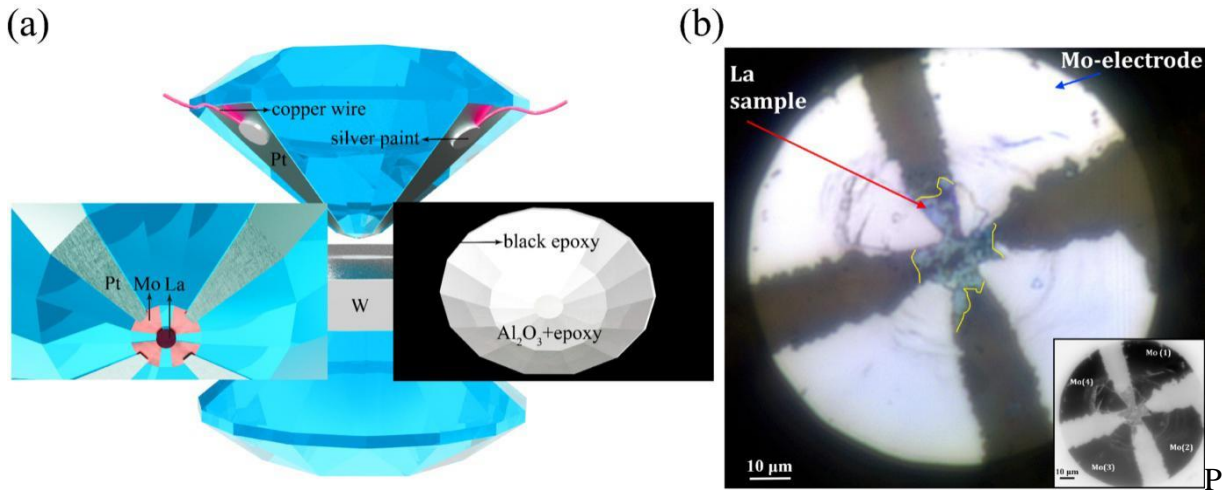

**FIG. 1.** (a) Schematic of the assembly used for the electrical resistance measurements. The sample chamber consisted of a tungsten outer gasket (W) with an insulating Al$_2$O$_3$ and epoxy. (b) Microphotograph of a sample at 140 GPa illuminated from the top. An inset shows the sample illuminated from the bottom.

The EoS of the metallic La has been investigated in the DAC and MgO as the PTM. The selected synchrotron XRD patterns at various pressures in the pressure range of 12-139 GPa are shown in Figure S1. By using the Rietveld refinement, the experimental XRD patterns of La sample are presented at 49 GPa, 74 GPa, 99 GPa and 136 GPa, respectively (Fig. 2a). The patterns are determined as dist. *fcc* (*R-3m*) and *fcc* (*Fm-3m*) phase at 49 GPa and 74 GPa. At 99 GPa and 136 GPa, Bragg peaks of La in the experimental patterns can be well indexed by a distorted *fcc*-La structure (with space group *Fmmm*), which hasn't been discovered in other rare earth metals before [27]. As shown in Fig. S2, to make sure this phase fits the experimental result best, we compared it with another two possible phases (*Fm-3m* and *R-3m*) at 123 GPa. Combined the analysis of XRD data and result of electrical measurements, we roughly define the phase boundary at around 80 GPa. The Le Bail refinements of the lattice parameters and volume of this distorted *fcc*-La structure at various pressures can be seen in Supporting Information Table S1. In the experimental Run 2, we also triggered this new phase, which can be seen from Supporting Information Figure S3 and Figure S4. The refined lattice parameters and unit cell volume of La upon the whole compression are shown as a function of pressure in Fig. 2b and Fig. 2c. The diagrams show the anisotropic behavior of the La sample under pressure: *da/dP* and *db/dP* are close and can be roughly approximated by linear functions $a(P) = 4.3316 - 0.0033 \cdot P$ (Å) and $b(P) = 4.3562 - 0.0032 \cdot P$ (Å), while *dc/dP* is different: $c(P) = 4.3461 - 0.0028 \cdot P$ (Å). In order to determine the parameters of the EoS, the obtained pressure-volume data were fitted by the third-order Birch-Murnaghan equation[28]:

$$P = \frac{3B_0}{2}\left[\left(\frac{V}{V_0}\right)^{\frac{-7}{3}} - \left(\frac{V}{V_0}\right)^{\frac{-5}{3}}\right]\left\{1 + \frac{3}{4}(B_0' - 4)\left[\left(\frac{V}{V_0}\right)^{\frac{-2}{3}} - 1\right]\right\}, \qquad (1)$$

namely $V_0$, $B_0$ and $B_0'$, where $V_0$ is the equilibrium cell volume, $B_0$ is the bulk modulus, and $B_0'$ is the derivative of bulk modulus with respect to pressure. The fitted parameters are $V_0 = 35.2\ (4)$ Å$^3$, $B_0 = 27\ (1)$ GPa and $B_0' = 4$. The present bulk modulus is consistent with previous measurement of $B_0 = 24$ GPa [29-30]. In Fig. 2b, it is clearly seen that the lattice parameters changed discontinuously with pressures, indicating two phase transitions. In contrast, the volume data didn't show clear discontinuities up to the highest pressure, which is also observed in similar rare earth metals [31-33]. This may be due to the fact that our observed dist. *fcc* phase (*R-3m*) and dist. *fcc* phase (*Fmmm*) represent small different stacking sequences of *fcc* phase (*R-3m*) and thus may obscure observation of any volume discontinuity in the transitions.

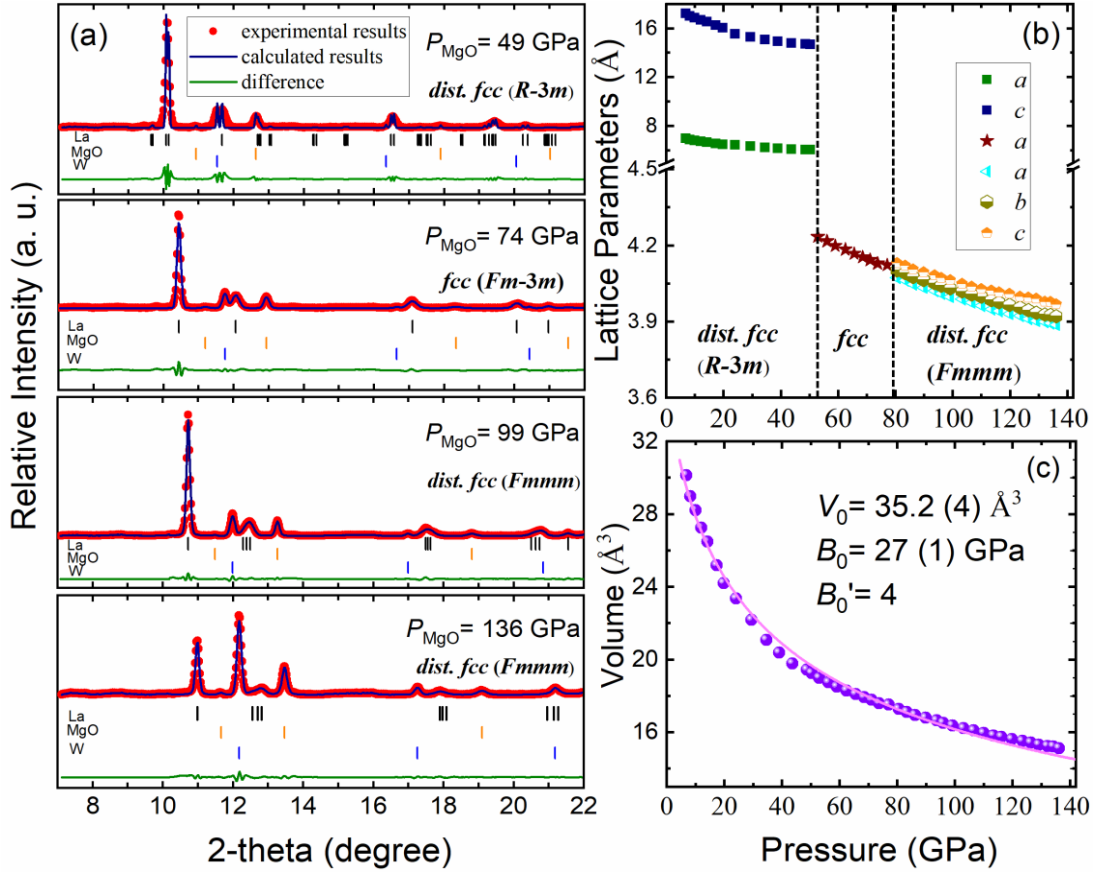

**FIG. 2.** (a) Rietveld refinement of the experimental XRD pattern of La sample at 49 GPa, 74 GPa, 99 GPa and 136 GPa, respectively. Positions of the Bragg reflections from La, MgO, and W gasket are marked by black, yellow and blue vertical ticks, respectively. (b) The lattice parameters as a function of pressure. The straight dash line marked the transition pressure points. (c) Pressure-volume relations for La sample. The curve shows the third-order Birch-Murnaghan equation fitting to the experimental data (solid spheres).

Three experimental runs of superconductivity of La were investigated in the DAC with a 100-μm or 50-μm culet and $Al_2O_3$ as the PTM. After loading La sample into the DAC (the detailed setup is shown in Fig. 1), we measured the evolution of its electrical resistance as a function of temperature at various pressures in experimental Run 1-3, as shown in Fig. 3 and Supporting Information Figure S5. Most of the superconducting transition critical temperature $T_c$ is defined near the temperature where the resistance begins to drop, the exact position depends on the situation of experiment. In Fig. 3, at ambient pressure, during the cooling from 300 K to 1.5 K, the resistance first displays a typical metal-like behavior, then, at 5.3 K, decreases sharply to zero. Our obtained superconducting $T_c$ at ambient pressure is in good agreement with previous work [15, 17]. Below 60 GPa, $T_c$ increase firstly with the increase pressure and then declines at 18 GPa, which shows the similar tendency with the reported data [15, 17]. Upon further compression, the evolution of the electrical resistance as a function of temperature at pressures above 60 GPa is also shown in Fig. 3. We found the superconducting transitions can be triggered up to the maximum pressure of 140 GPa, and the superconducting $T_c$ has been falling down to 2.2 K at 140 GPa. The superconducting transitions were obvious, with the transition width around 0.5 K, (from 10% to 90% of the normal state resistance at $T_{onset}$), indicating a good homogeneity of the superconducting phase. The beginning of the transition to a superconducting state at 140 GPa can also be fixed at the initial point of the resistance growth (~ 4.8 K), which is caused by the presence of distributed superconducting-normal grain boundaries [34].

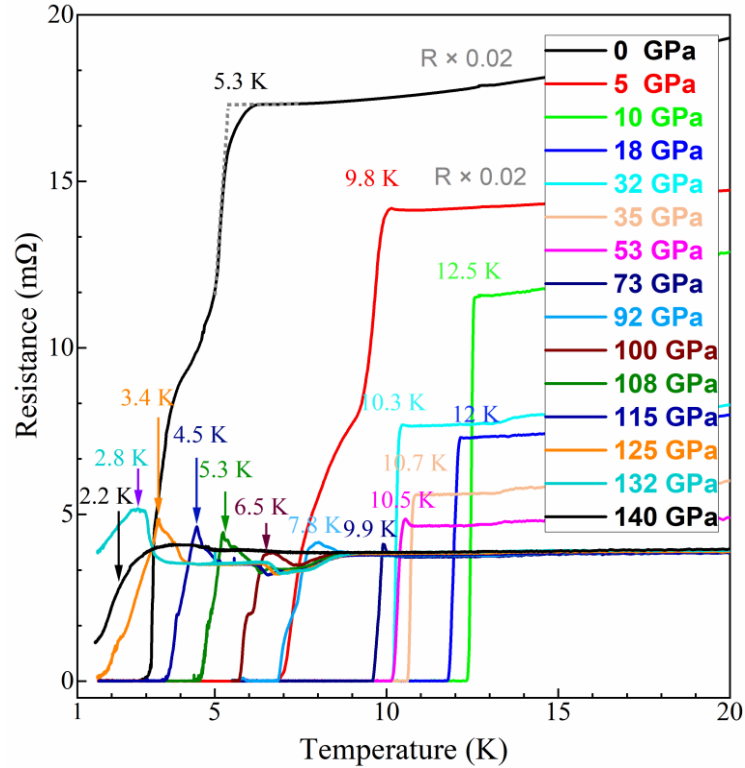

**FIG. 3.** Resistance of the La sample as a function of temperature at various pressures during cooling in the experimental Run 1. The gray characters represent the magnification of R. The dotted gray line and arrows show the definition of the superconducting transition temperature $T_c$.

To get a further proof that the phenomenon observed (Fig. 3 and Supporting Information Figure S5) is indeed the superconducting transition, we measured the electrical resistance around the transition temperature in various external magnetic fields at 53 GPa (Supporting Information Figure S6) and 140 GPa. Fig. 4a shows the measured resistance at 140 GPa with applied magnetic fields $H$ of 0.025, 0.05, 0.075 and 0.1 T. The critical temperature $T_c$ decreased with increasing magnetic field, indicating that the transition is superconducting. To suppress superconductivity, much stronger fields are required. To estimate the upper critical magnetic field $H_{c2}(0)$, we applied an extrapolation method combined with Ginzburg-Landau (GL) equation [35]:

$$\mu_0 H_{C2}(T) = \mu_0 H_{C2}(0)\left(1 - \frac{T^2}{T_C^2}\right) \quad (2)$$

The extrapolation ($R^2 = 0.98$) of the transition temperature gives an estimate of $\mu_0 H_{c2}(0) = 0.32$ T. The Werthamer-Helfand-Hohenberg (WHH) model [36] for the critical magnetic field, simplified by Baumgartner [37],

$$\mu_0 H_{C2}(T) = \frac{\mu_0 H_{C2}(0)}{0.693}\left[\left(1 - \frac{T}{T_C}\right) - 0.153 \cdot \left(1 - \frac{T}{T_C}\right)^2 - 0.152 \cdot \left(1 - \frac{T}{T_C}\right)^4\right] \quad (3)$$

leads to $\mu_0 H_{c2}(0) = 0.43$ T.

To obtain further insight, we have calculated the superconducting parameters of La at various pressures. To compare our calculations with the experimental results at 50 GPa, we first computed the superconducting parameters of the slightly distorted *fcc*-La (*R-3m*) using the PZ and PBE pseudopotentials, obtaining $\lambda = 1.07$, $\omega_{\log} = 113$ K, and $T_c = 9.3$ K at $\mu^* = 0.1$ (50 GPa). The Eliashberg functions $\alpha^2 F(\omega)$ of the distorted *fcc*-La with different σ-broadening (QE) at 50 GPa are presented in Supporting Information Figure S7. The reported experimental $T_c$ of 10.5 K at 50 GPa [17] is consistent with our experimental results (53 GPa, 10.5 K) and close to the calculated value (9.3 K). Thus, we are confident that our theoretical calculations of $T_c$ and $B_{c2}$ as a function of pressure are reliable. At 130 GPa, we took into account that the phonon spectrum of La ends at 350 cm$^{-1}$ (~ 0.043 eV, Fig. 4b). At such energies, the density of states near $E_F \pm \hbar\omega_{\max}$ is almost constant (≈ 10.05 states/Ry/La), which allows us to apply the "constant DOS approximation" [38] and take $\alpha^2 F(\omega)$ corresponding to an almost zero broadening in the QE output. The results obtained with the Goedecker-Hartwigsen-Hutter-Teter PZ and PAW PBE pseudopotentials are the same: $\lambda = 0.69$, $\omega_{\log} = 288$ K, $T_c$ (A-D) = 10.1 K for $\mu^* = 0.1$ at 130 GPa. There is no significant difference in $T_c$ and $\mu^*$ obtained with the different pseudopotentials. The experimental critical temperature of 2.8 K corresponds to an anomalous value of $\mu^* = 0.21$. At this Coloumb pseudopotential, the McMillan isotope coefficient $\beta = 0.20$ is quite small (it increases to 0.37 at $\mu^* = 0.15$), while the coherence length $\xi_{BCS} = 0.5\sqrt{h/\pi e\, H_{C2}}$ is 23 nm, which is about 35% lower than for the metallic lanthanum at 0 GPa (36 nm) [39].

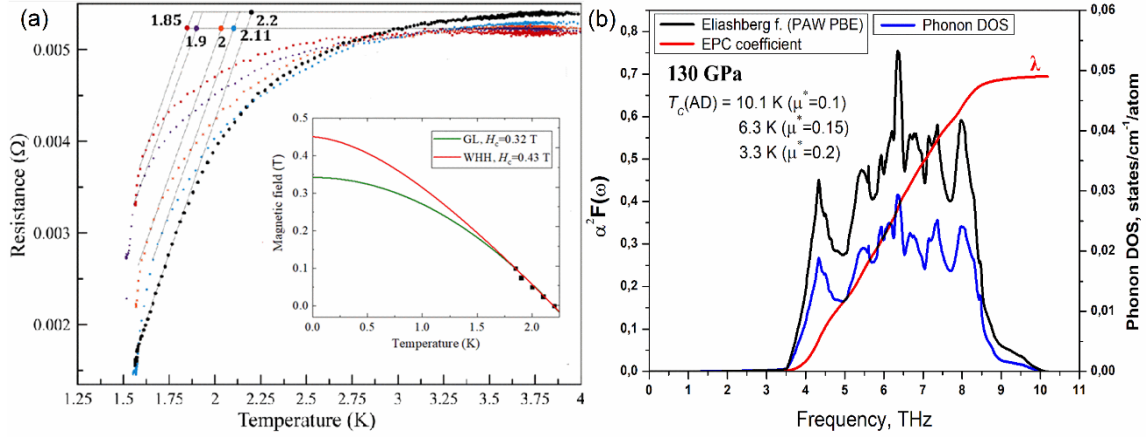

**FIG. 4.** (a) The experimental measured superconducting transition of the La sample at 140 GPa with a magnetic field applied. In an inset, the dots show the measured values of $T_c$ and magnetic field, while the solid lines represent fittings by the WHH and LG equations. (b) The calculated superconducting parameters of the distorted *fcc*-La phase at 130 GPa.

Fig. 5 shows the evolution of the crystal structure and superconducting transition temperature as a function of pressure in the form of a phase diagram. At about 80 GPa, $T_c$ decreases faster during the compression which reveals the start of distortion of cell parameters. The pressure dependence of the superconducting $T_c$ of the metallic La exhibits an approximately linear behavior above 80 GPa. For comparison, we plotted the data obtained in this study and earlier results together. The differences between the $T_c$ values reported in this work and those from the earlier studies [15, 17] are very small and can be attributed to different hydrostatic conditions of various PTM. For the electrical conductivity measurements, we used an excessive amount of $Al_2O_3$ as the PTM. As an example, we investigated the pressure distribution in the superconducting electrical resistance measurements at about 132 GPa, as shown in Supporting Information Figure S8. The largest pressure difference of 7 GPa was 20 microns away from the sample's center. Considering that MgO solid is softer than $Al_2O_3$ and probably offering a better hydrostatic environment at ultrahigh pressures, we presumed that the pressure difference in the XRD experiments was no more than 7 GPa. Thus the presented experiments show good quasi-hydrostatic pressure conditions.

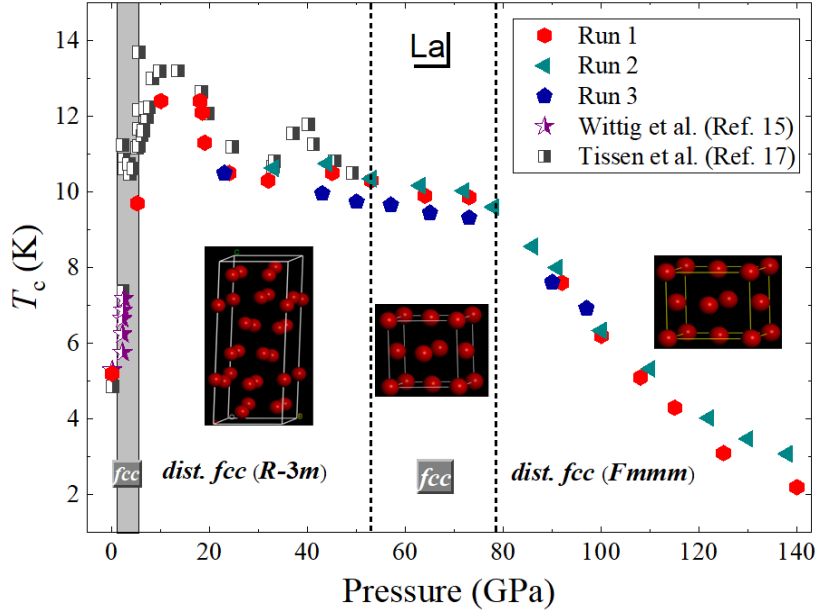

**FIG. 5.** The phase diagram and superconducting transition temperature $T_c$ of La as a function of pressure. The red hexagon, green triangle and blue pentagon symbols denote our experimental superconducting $T_c$ data from Run 1-3. The purple pentagrams and black squares represent the data from Wittig *et al.* [15] and Tissen *et al.* [17]. The black dashed line marks the phase transition points determined by the present study.

## Conclusions

In conclusion, we have studied the crystal structure of the metallic La at pressures up to 140 GPa and discovered a previous unknown distorted *fcc*-La phase (space group *Fmmm*). The superconducting transitions have been detected in the distorted *fcc*-La at 9.6 K (78 GPa) and decreases to at 2.2 K (140 GPa) by means of the four-probe resistance measurements. We have investigated the influence of an external magnetic field (0-0.1 T) on $T_c$ at 140 GPa which allowed us to estimate the upper critical magnetic field $\mu_0H_{c2}(0)$ at 0.32 T-0.43 T, according to the GL and WHH models, respectively. The calculations of the electron-phonon interactions within the classical BCS mechanism show a good stability and point to an increased Coloumb pseudopotential $\mu^*$=0.21, while the calculations using the regular $\mu^*$ interval (0.1-0.15) lead to an overestimated $T_c$ (6.3-10.1 K at 130 GPa). The superconducting properties of the metallic lanthanum exhibit an approximately linear behavior in the 78-140 GPa pressure range, and experimental $T_c$ = 3.3 K at 130 GPa and upper critical magnetic field $\mu_0H_{c2}(0)$ = 0.4 T at 140 GPa, are quite close to the result of calculations within the classical electron-phonon pairing mechanism at Coulomb pseudopotential $\mu^*$=0.21.

## Acknowledgments

This work was supported by the National Key R&D Program of China (grant no. 2018YFA0305900), National Natural Science Foundation of China (grant nos. 51632002, 11974133 and 51720105007), National Key Research and Development Program of China (grant no. 2016YFB0201204), Program for Changjiang Scholars and Innovative Research Team in University (grant no. IRT_15R23), and National Fund for Fostering Talents of Basic Science (grant no. J1103202). The calculations were performed on the Rurik supercomputer at the MIPT and the Arkuda supercomputer of the Skolkovo Foundation. D.V . S. thanks the Russian Foundation for Basic Research (RFBR) project №19-03-00100.

# SUPPORTING INFORMATION

# Superconductivity and Equation of State of Lanthanum at Megabar Pressures

Wuhao Chen[1], Dmitrii V. Semenok[2], Ivan A. Troyan[3], Anna G. Ivanova[3], Xiaoli Huang[1,*], Artem R. Oganov[2,4] and Tian Cui[5,1,*]

[1] State Key Laboratory of Superhard Materials, College of Physics, Jilin University, Changchun 130012, China

[2] Skolkovo Institute of Science and Technology, Skolkovo Innovation Center, 3 Nobel Street, Moscow 143026, Russia

[3] Shubnikov Institute of Crystallography, Federal Research Center Crystallography and Photonics, Russian Academy of Sciences, Moscow 119333, Russia

[4] International Center for Materials Discovery, Northwestern Polytechnical University, Xian 710072, China

[5] School of Physical Science and Technology, Ningbo University, Ningbo 315211, China

*Corresponding authors: huangxiaoli@jlu.edu.cn and cuitian@jlu.edu.cn

# CONTENT

# 1. Equations for calculating $T_c$ and related parameters

The superconducting transition temperature $T_C$ was estimated by using the Allen-Dynes formula:

$$T_c = \omega_{log} \frac{f_1 f_2}{1.2} \exp\left(\frac{-1.04(1+\lambda)}{\lambda - \mu^* - 0.62\lambda\mu^*}\right) \tag{S1}$$

where the product of the Allen-Dynes coefficient is:

$$f_1 f_2 = \sqrt[3]{1 + \left(\frac{\lambda}{2.46(1+3.8\mu^*)}\right)^{\frac{3}{2}}} \cdot \left(1 - \frac{\lambda^2(1-\omega_2/\omega_{log})}{\lambda^2 + 3.312(1+6.3\mu^*)^2}\right) \tag{S2}$$

The EPC constant $\lambda$, logarithmic average frequency $\omega_{log}$ and the mean square frequency were calculated as:

$$\lambda = 2 \int_0^{\omega_{max}} \frac{\alpha^2 F(\omega)}{\omega} d\omega \tag{S1}$$

$$\omega_{log} = e^{\left(\frac{2}{\lambda}\int_0^{\omega_{max}} \frac{d\omega}{\omega} \alpha^2 F(\omega) \ln(\omega)\right)}, \quad \omega_2 = \sqrt{\frac{1}{\lambda}\int_0^{\omega_{max}} \left[\frac{2\alpha^2 F(\omega)}{\omega}\right] \omega^2 d\omega} \tag{S2}$$

To calculate the isotopic coefficient $\beta$, the Allen-Dynes interpolation formulas were used:

$$\beta_{McM} = -\frac{d \ln T_C}{d \ln M} = \frac{1}{2}\left[1 - \frac{1.04(1+\lambda)(1+0.62\lambda)}{[\lambda - \mu^*(1+0.62\lambda)]^2} \mu^{*2}\right] \tag{S5}$$

$$\beta_{AD} = \beta_{McM} - \frac{2.34\mu^{*2}\lambda^{3/2}}{(2.46+9.25\mu^*) \cdot ((2.46+9.25\mu^*)^{3/2} + \lambda^{3/2})} -$$

$$- \frac{130.4 \cdot \mu^{*2}\lambda^2 (1+6.3\mu^*)\left(1 - \frac{\omega_{log}}{\omega_2}\right)\frac{\omega_{log}}{\omega_2}}{\left(8.28 + 104\mu^* + 329\mu^{*2} + 2.5 \cdot \lambda^2 \frac{\omega_{log}}{\omega_2}\right) \cdot \left(8.28 + 104\mu^* + 329\mu^{*2} + 2.5 \cdot \lambda^2 \left(\frac{\omega_{log}}{\omega_2}\right)^2\right)} \tag{S6}$$

where the last two correction terms are usually small (~0.01).
The Sommerfeld constant was found as

$$\gamma = \frac{2}{3}\pi^2 k_B^2 N(0)(1+\lambda), \tag{S7}$$

and was used to estimate the upper critical magnetic field and the superconductive gap by the well-known semi-empirical equations of the BCS theory (see J. P. Carbotte, *Rev. Mod. Phys.*, 62(4), 1990, equations 4.1 and 5.11), working satisfactorily for $T_C/\omega_{log} < 0.25$:

$$\frac{\gamma T_C^2}{(\mu_0 H_{C_2}(0))^2} = 0.168\left[1 - 12.2\left(\frac{T_C}{\omega_{log}}\right)^2 \ln\left(\frac{\omega_{log}}{3T_C}\right)\right] \tag{S8}$$

$$\frac{2\Delta(0)}{k_B T_C} = 3.53\left[1 + 12.5\left(\frac{T_C}{\omega_{log}}\right)^2 \ln\left(\frac{\omega_{log}}{2T_C}\right)\right] \tag{S9}$$

## 2. Structural information and results of X-ray diffraction studies

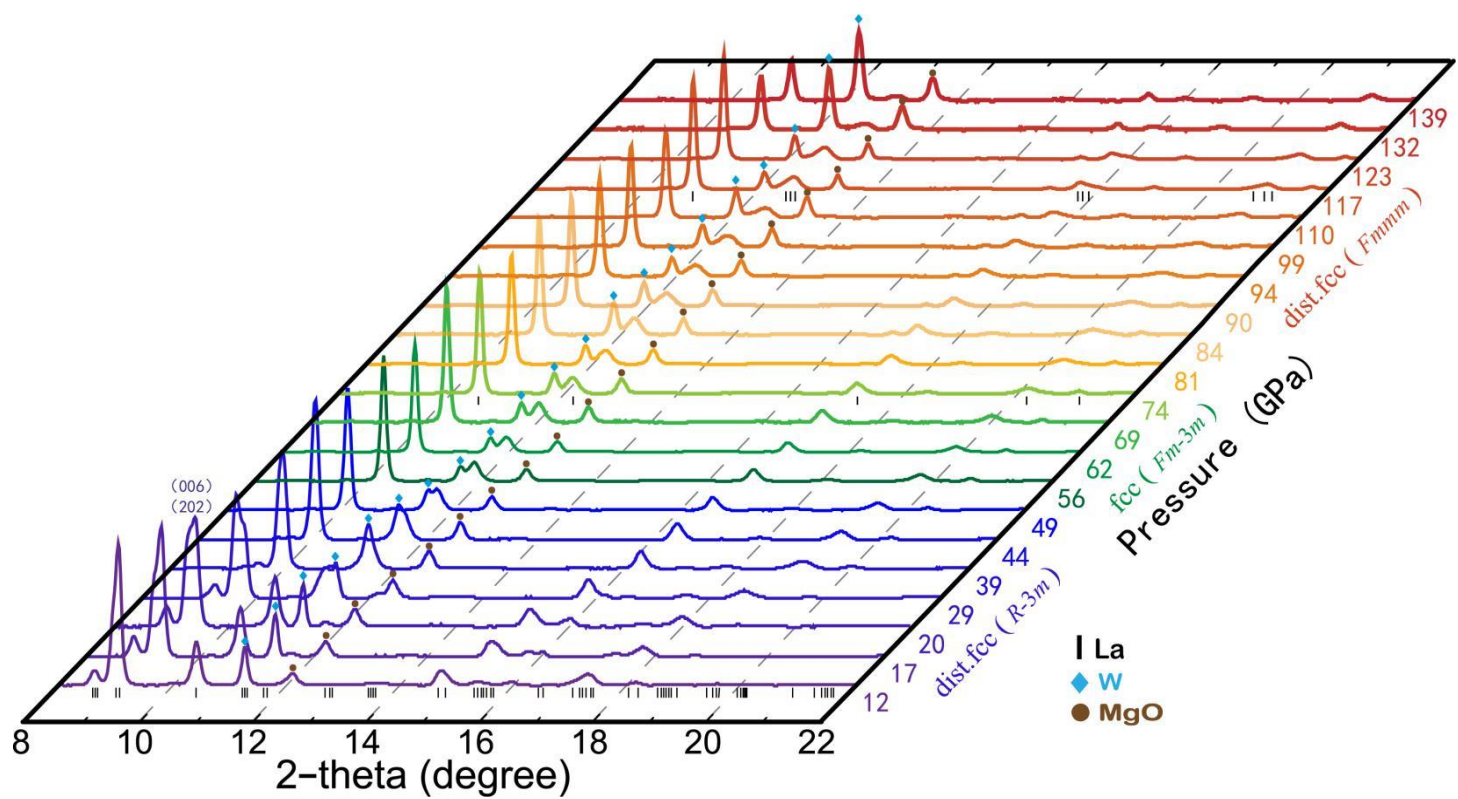

**Figure S1.** The selected synchrotron XRD patterns at various pressures in the pressure range of 12-139 GPa. The wavelength of the incident X-ray is 0.434 Å. Positions of the Bragg reflections from La sample, W gasket and MgO are marked with individual symbols.

**Table S1.** The experimental cell parameters for the orthorhombically distorted *fcc*-La (*Fmmm*) phase.

| Pressure, GPa | *a*, Å | *b*, Å | *c*, Å | *V*, Å³/atom |
|:---:|:---:|:---:|:---:|:---:|
| 81 | 4.0735 | 4.1025 | 4.1328 | 17.275 |
| 83 | 4.0628 | 4.0918 | 4.1223 | 17.125 |
| 86 | 4.0473 | 4.0768 | 4.1080 | 16.950 |
| 90 | 4.0347 | 4.0648 | 4.0963 | 16.800 |
| 94 | 4.0214 | 4.0520 | 4.0847 | 16.650 |
| 96 | 4.0113 | 4.0420 | 4.0740 | 16.525 |
| 99 | 3.9986 | 4.0318 | 4.0653 | 16.375 |
| 103 | 3.9889 | 4.0188 | 4.0505 | 16.225 |
| 110 | 3.9630 | 3.9980 | 4.0300 | 16.100 |
| 107 | 3.9758 | 4.0074 | 4.0400 | 15.960 |
| 113 | 3.9490 | 3.9850 | 4.0237 | 15.825 |

| 116 | 3.9427 | 3.9770 | 4.0171 | 15.750 |
| 120 | 3.9310 | 3.9660 | 4.0115 | 15.625 |
| 123 | 3.9233 | 3.9564 | 4.0003 | 15.525 |
| 126 | 3.9112 | 3.9453 | 3.9969 | 15.425 |
| 129 | 3.9034 | 3.9356 | 3.9921 | 15.325 |
| 132 | 3.8997 | 3.9289 | 3.9800 | 15.250 |
| 134 | 3.8944 | 3.9268 | 3.9797 | 15.225 |
| 136 | 3.8885 | 3.9219 | 3.9697 | 15.125 |

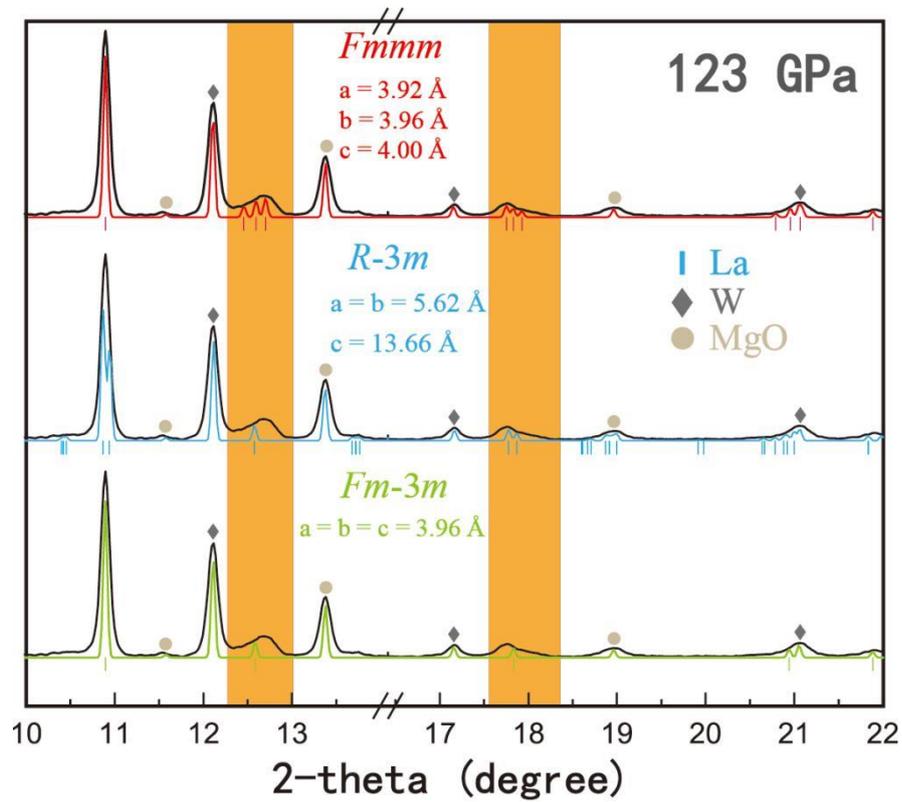

**Figure S2.** Comparison of the refinements of La experimental XRD patterns at 123 GPa with different phases. Black curves represent the same experimental data while the calculated results of *Fm*-3*m*, *R*-3*m* and *Fmmm* phases are shown in green, blue and red, respectively. Orange rectangles highlight the main differences among three phases.

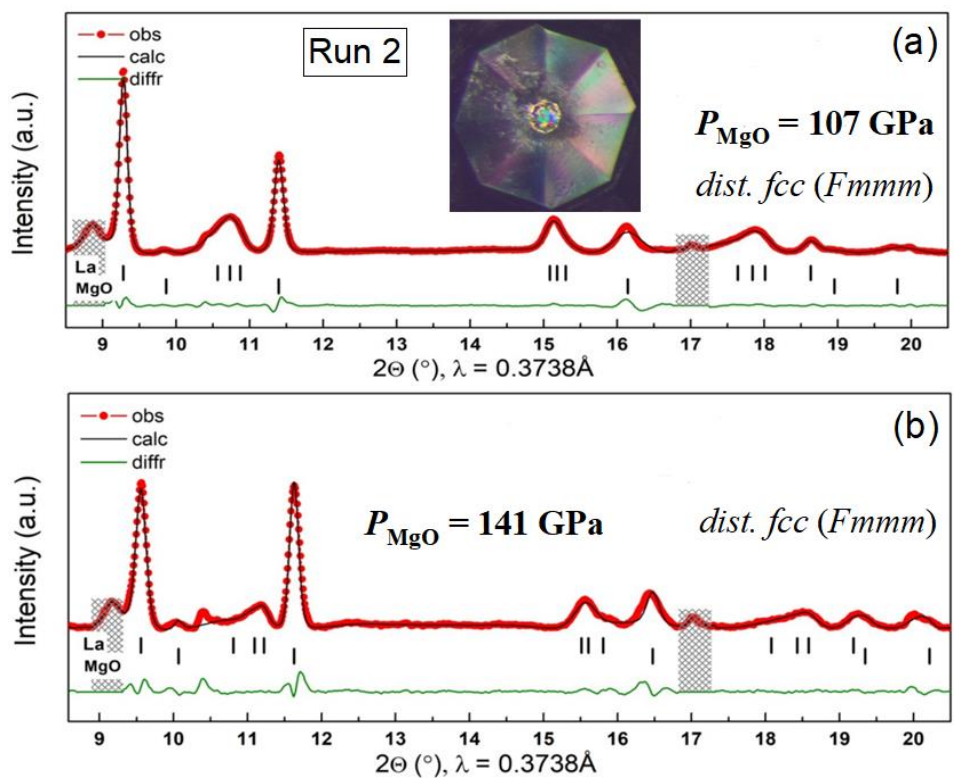

**Figure S3.** The Le Bail refinement of the experimental XRD patterns of La sample at various pressures in experimental Run 2. Side reflections were excluded from the analysis. Inset: a photo of the DAC with a 50 μm culet and La sample used in the X-ray diffraction studies at ESRF.

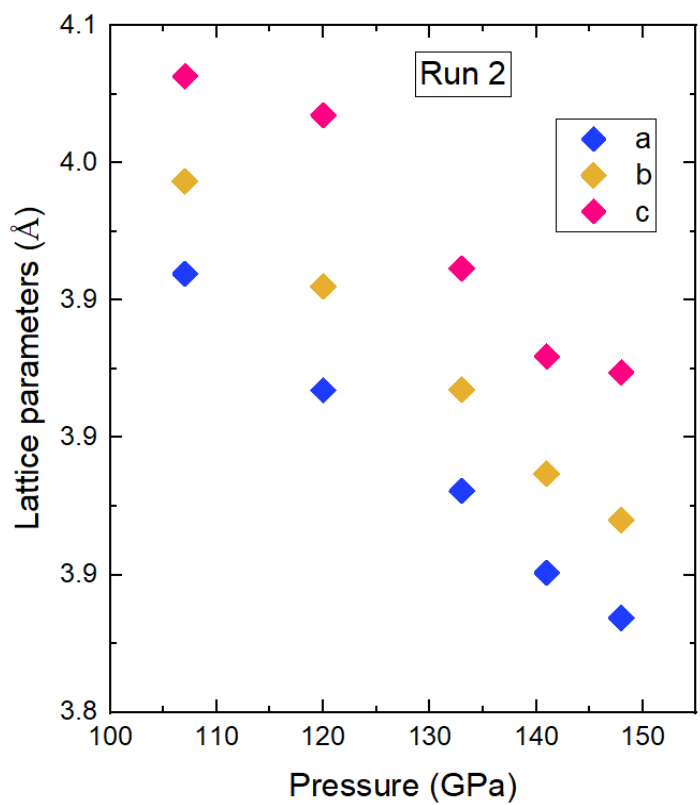

**Figure S4.** The refined cell parameters of the La sample at various pressures in experimental Run 2.

## 3. Superconducting properties

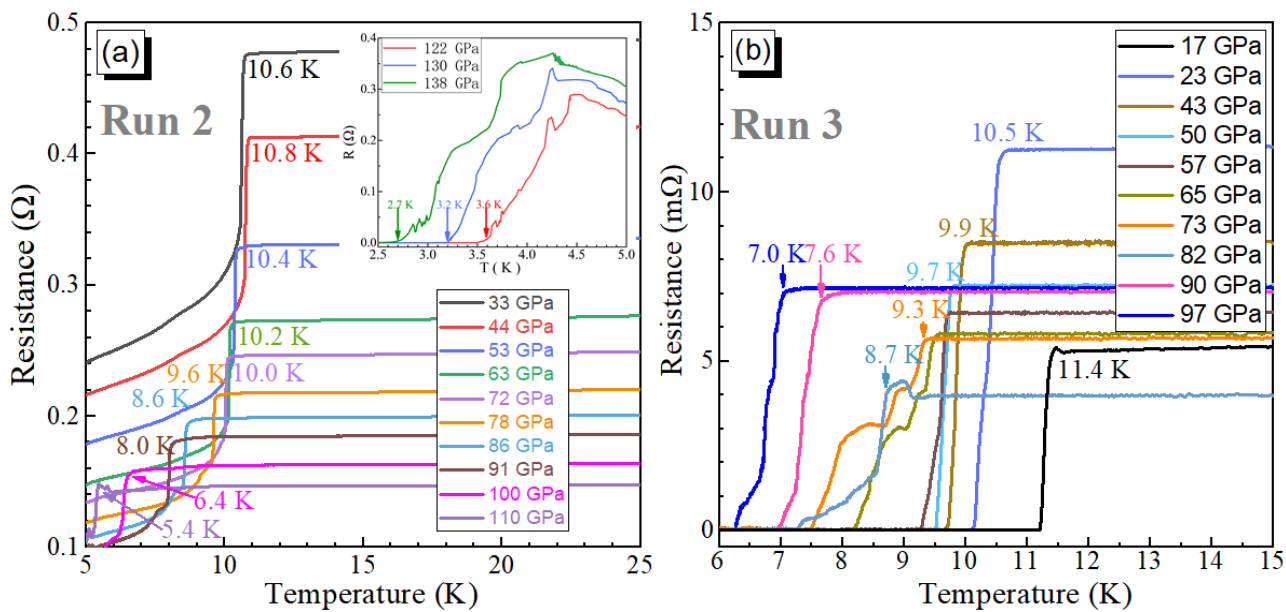

**Figure S5.** Experimental measured R(T) dependence of sample La at different pressures. (a) R(T) data of experimental Run 2 in the pressure range of 33-138 GPa, and (b) R(T) data of experimental Run 3 in the pressure range of 17-97 GPa.

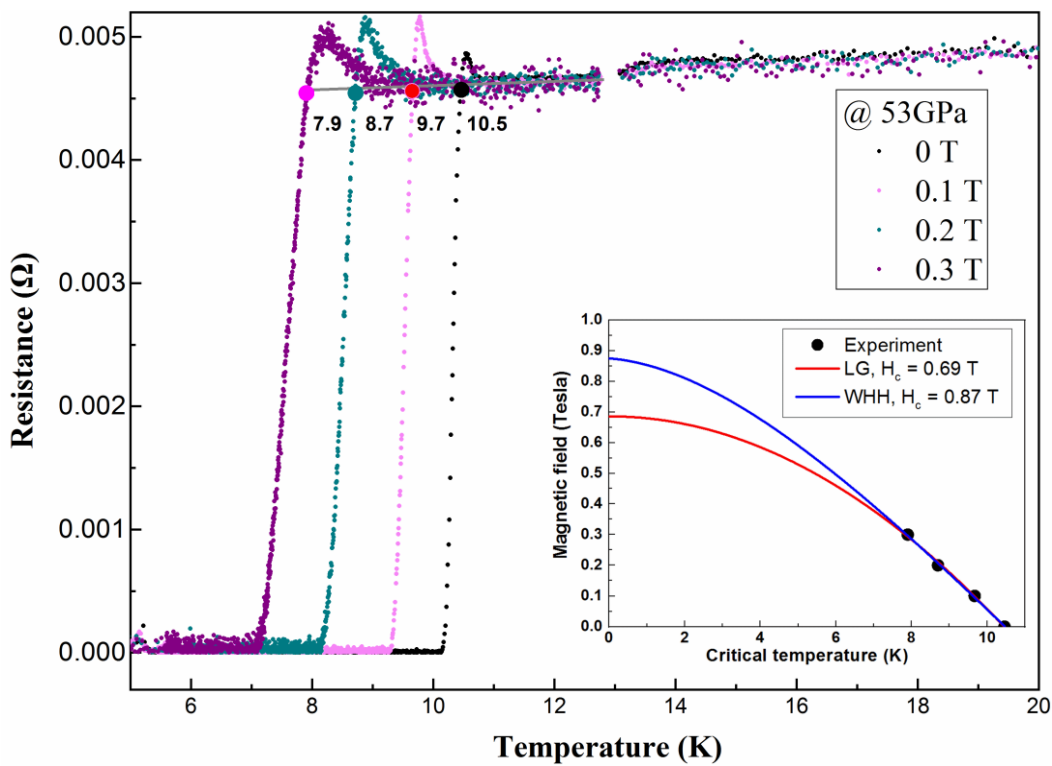

**Figure S6.** The superconducting transition of the La sample at 53 GPa with a magnetic field applied. The inset shows the change of $T_c$ with magnetic field, where the red and blue lines represent fittings by the LG and WHH equations, respectively.

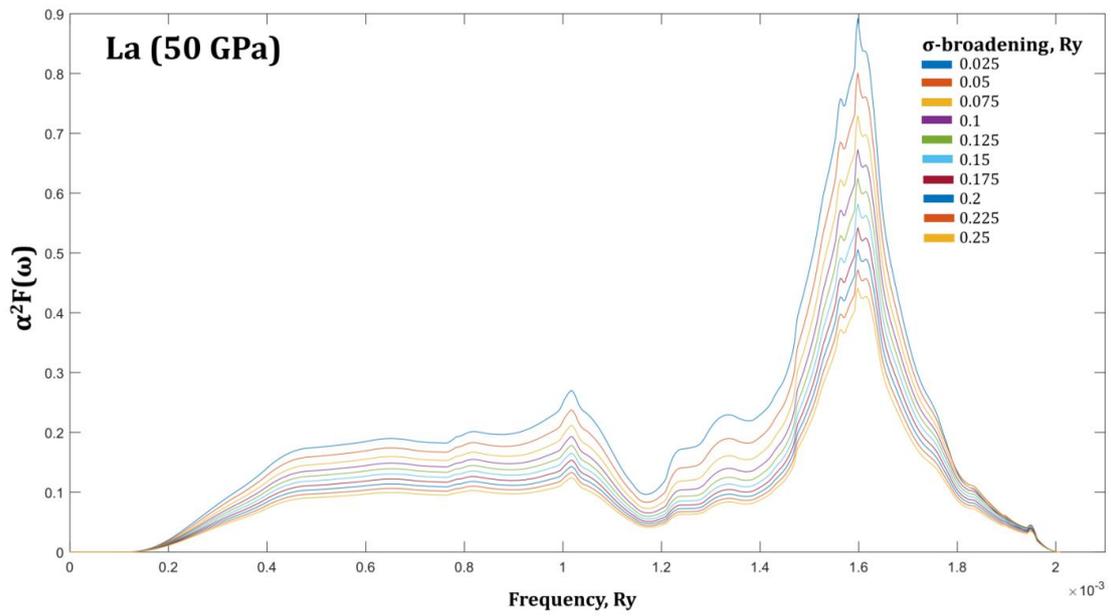

**Figure S7.** The Eliashberg functions $α^2F(ω)$ of the distorted *fcc*-La phase with different σ-broadening (QE) at 50 GPa.

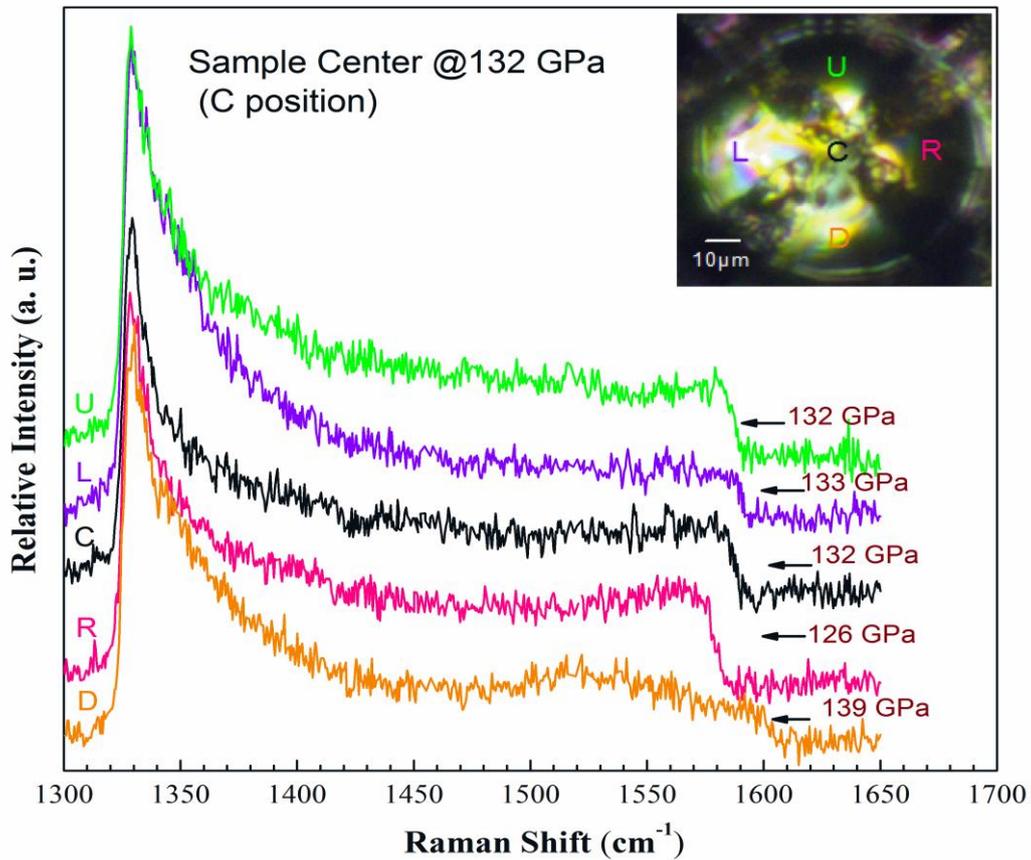

**Figure S8.** The pressure distribution in the superconducting electrical resistance measurements at about 132 GPa. The inset shows the corresponding sample positions for pressure measurements.